\documentclass[11pt]{article}
\usepackage{moriond,epsfig}

\def\be{\begin{equation}}
\def\ee{\end{equation}}
\def\bea{\begin{eqnarray}}
\def\eea{\end{eqnarray}}

\begin{document}  
\hfill \parbox{4cm}{CERN-TH/2002-99 \\ TTP02-05}
\vspace*{3.5cm}
\title{RADIATIVE RETURN AT NLO: \\ THE PHOKHARA MONTE CARLO GENERATOR
\footnote{Work supported in part by BMBF under grant number 05HT9VKB0 and 
E.U. EURO\-DA$\Phi$NE network TMR project FMRX-CT98-0169.}}

\author{Germ\'an Rodrigo~$^{1,}$
\footnote{Supported in part by E.U. TMR grant HPMF-CT-2000-00989; 
e-mail: {\tt german.rodrigo@cern.ch}},
Henryk Czy\.z~$^{2,}$
\footnote{Supported in part by EC 5-th Framework, 
contract HPRN-CT-2000-00149; e-mail: {\tt czyz@us.edu.pl}} 
and Johann H. K\"uhn~$^{1,3,}$
\footnote{e-mail: {\tt jk@particle.uni-karlsruhe.de}} }

\address{${}^1$ Theory Division, CERN, CH-1211 Geneva 23, Switzerland. \\ 
${}^2$ Institute of Physics, University of Silesia,
PL-40007 Katowice, Poland. \\
${}^3$ Institut f\"ur Theoretische Teilchenphysik,
Universit\"at Karlsruhe, D-76128 Karlsruhe, Germany.}

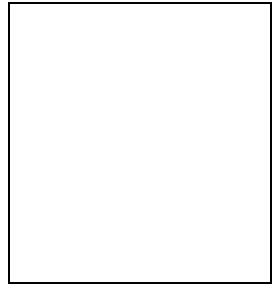
\maketitle\abstracts{
Electron--positron annihilation into hadrons plus an 
energetic photon from initial state radiation allows the hadronic 
cross-section to be measured over a wide range of energies
at high luminosity meson factories.
A Monte Carlo generator called PHOKHARA has been constructed, 
which simulates this process at the next-to-leading order accuracy.}

The total cross section for electron--positron annihilation into
hadrons is one of the fundamental observables in particle physics. Its
high energy behaviour provides one of the first and still most
convincing arguments for the point-like nature of quarks. Its
normalization was evidence for the existence of quarks of three
different colours, and the recent, precise measurements even allow
for an excellent determination of the strong coupling at very high 
and intermediate energies through the influence of QCD corrections.

Weighted integrals over the cross section with properly chosen kernels
are, furthermore, a decisive input for electroweak precision tests.
This applies, for example, to the electromagnetic coupling at higher
energies or to the anomalous magnetic moment of the muon.
Of particular importance for these two applications is the low energy region,
say from threshold up to centre-of-mass (cms) energies of approximately 
3~GeV and 10~GeV, respectively. Recent measurements based on energy scans 
between 2 and 5~GeV have improved the accuracy in part of this
range. However, similar, or even further improvements below 2~GeV
would be highly welcome. The region between 1.4~GeV and 2~GeV, in
particular, is poorly studied and no collider will cover this region 
in the near future. Improvements or even an independent cross-check 
of the precise measurements of the pion form factor in the low 
energy region by the CMD2 and DM2 collaborations 
would be extremely useful, since this dominates in
the analysis of the muon anomalous magnetic moment.

Experiments at present electron--positron colliders operate mostly at 
fixed energies, albeit with enormous luminosity, with BaBar
and BELLE at 10.6~GeV and KLOE at 1.02 GeV as most prominent 
examples. This peculiar feature allows the use of the radiative return, 
i.e. the reaction
\begin{equation}
e^+(p_1)+ e^- (p_2) \to \gamma(k) + \gamma^*(Q) (\to \mathrm{hadrons})~,
\label{eq:reaction}
\end{equation}
to explore a wide range of $Q^2$ in a single 
experiment~\cite{Binner:1999bt,Czyz:2000wh}.
Nominally masses of the hadronic system between $2 m_\pi$ and the cms
energy of the experiment are accessible. In practice it is useful to
consider only events with a hard photon --- tagged or untagged ---
to clearly identify  the reaction, which lowers the energy 
significantly.

The study  of events with photons emitted under both large and small 
angles, and thus at a significantly enhanced rate, is particularly 
attractive for the $\pi^+\pi^-$ final state with its clear signature, 
an investigation performed at present at DA$\Phi$NE~\cite{kloe}.
Events with a tagged photon, emitted under a large angle with
respect to the beam, have a clear signature and are thus particularly
suited to the analysis of hadronic final states of higher
multiplicity~\cite{babar}.

The inclusion of radiative corrections is essential for the precise
extraction of the cross section, which is necessarily based on a Monte
Carlo simulation. A first program, called EVA, was constructed
some time ago~\cite{Binner:1999bt}. It simulates $\pi^+\pi^-$ events
and includes initial-state radiation (ISR), final-state radiation (FSR), 
and their interference at the leading order (LO), and the dominant 
radiative corrections from additional collinear radiation through 
structure function techniques. Four pion production 
were also considered~\cite{Czyz:2000wh}.

The complete NLO corrections have been implemented 
in a new Monte Carlo generator called PHOKHARA~\cite{Rodrigo:2001kf}. 
This includes virtual and soft photon corrections~\cite{Rodrigo:2001jr} 
to the process $e^+ e^- \rightarrow \gamma+\gamma^*$ and the emission of 
two real hard photons: $e^+ e^- \rightarrow \gamma+\gamma+\gamma^*$.
The current version includes ISR only and is limited to 
$\pi^+ \pi^- \gamma (\gamma)$ and $\mu^+ \mu^- \gamma (\gamma)$
as final states. The uncertainty from unaccounted higher order
ISR was estimated~\cite{Rodrigo:2001kf} at around $0.5\%$. 
The dominant FSR contribution can be deduced from the earlier program 
EVA and will be implemented in the forthcoming versions of PHOKHARA.
Both programs were initially designed to simulate reactions with 
tagged photons, i.e. at least one photon was required to be emitted 
under large angles. The extension of these results to untagged 
photon events, i.e for photons emitted at arbitrary small angles,
has been recently investigated~\cite{Rodrigo:2001cc,Kuhn:2002xg}.
Due to the modular structure of PHOKHARA additional hadronic modes can be 
easily implemented. The modes with three and four pions are in preparation. 
Further information can be obtained at
{\tt http://cern.ch/german.rodrigo/phokhara}.

\end{document}